# Relation between static short-range order and dynamic heterogeneities in a nanoconfined liquid crystal

Ronan Lefort,[1] Denis Morineau,[1] Régis Guégan,[2] Mohammed Guendouz,[3] Jean-Marc Zanotti,[4] and Bernhard Frick[5]

[1]*Institut de Physique de Rennes (IPR), CNRS-UMR 6251, Université de Rennes 1, 35042 Rennes, France*
[2]*Institut des Sciences de la Terre (ISTO), 1A, rue de la Férollerie, 45071 Orléans cedex 2, France*
[3]*Laboratoire d'Optronique, FOTON, CNRS-UMR 6082, Université de Rennes 1, 22302 Lannion, France*
[4]*Laboratoire Léon Brillouin, CEA-Saclay, 61191 Gif-Sur-Yvette, France*
[5]*Institut Laue-Langevin, 6 rue Jules Horowitz, 38042 Grenoble, France*

We analyze the molecular dynamics heterogeneity of the liquid crystal 4-*n*-octyl-4-cyanobiphenyl nanoconfined in porous silicon. We show that the temperature dependence of the dynamic correlation length $\xi_{wall}$, which measures the distance over which a memory of the interfacial slowing down of the molecular dynamics persists, is closely related to the growth of the short-range static order arising from quenched random fields. More generally, this result may also shed some light on the connection between static and dynamic heterogeneities in a wide class of condensed and soft matter systems.

The concept of dynamic heterogeneity plays a key role in the current understanding of the dynamic properties of systems as diverse as polymers, supercooled liquids, colloidal suspensions, or granular media [1–4]. All of these systems have in common a tremendous slowing down of their structural relaxation when approaching their glass or jamming transition. The considerable experimental and theoretical interest in this matter has led to the general agreement that the slowing down should be driven by a substantial increase of some characteristic length. Very recently, the introduction of multipoint dynamical susceptibilities conciliating theoretical views on the glass transition and established experimental data [5,6] have further enlightened the intrinsic dynamic character of this correlation length (denoted *length of dynamic cooperativity*). Also available models were extended to naturally heterogeneous systems like fluids confined in nanoporous solids [7,8]. Although it may not be ubiquitous to all glass-forming systems, the static nature of the observed heterogeneity is still debated [9]. For supercooled molecular liquids, some authors have invoked the existence of an underlying local structural order despite the lack of any direct signature in the experimental structure factor [10]. On the other hand, there exist some more complex fluids that can naturally develop static order parameters, and one may legitimately wonder in that case whether this static order couples to dynamic heterogeneities. Liquid crystals (LCs) constitute archetype examples of such systems, which experience nematic or smectic orders. The spatial correlation length $\xi_{stat}$ of the critical fluctuations of these order parameters is directly accessible to the experiment. Although the molecular dynamics of LCs is generally described by a finite set of correlation times and a well-defined relaxation mechanism [11–13], they also share many features with supercooled liquids on a time scale where collective critical fluctuations remain very slow [14]. Interestingly enough, this similarity between LCs and supercooled liquids can be emphasized by confining the LC in a nanoporous solid. Indeed, the broadening of relaxation time distributions is a salient feature of the molecular dynamics of confined systems [15–20]. It shows that interfacial effects do propagate into the pore volume over a distance comparable to the pore size itself. This motivates the introduction of a characteristic length $\xi_{wall}$ in the confined system, which defines the typical distance from the pore wall at which a molecular correlation time recovers its bulk value [15–17]. $\xi_{wall}$ provides therefore a measure of the dynamic heterogeneity in the confined fluid introduced by the solidliquid interface. This interface has also a very important influence on the structure of the confined phase: the interfacial energy can become comparable to the volume enthalpy and can considerably depress phase transition temperatures (Gibbs-Thompson effect). Topological effects (finite size, quenched disorder) will also prevent any correlation length of the fluid to grow larger than some fraction of the pore size. In the case of nanoconfined LCs, these lengths can correspond to nematic or smectic orders [21–25], providing therefore a unique opportunity to analyze how a frustrated static correlation length compares to that of inhomogeneous molecular dynamics. The aim of this Rapid Communication is to report on such a relation between $\xi_{wall}$ and a short-range smectic ordering, choosing the liquid crystal 4-*n*-octyl-4'-cyanobiphenyl (8CB) confined in porous silicon as a representative system.

Columnar porous silicon (PSi) matrices were obtained from a heavily doped (100)-oriented p-type silicon substrate by electrochemical etching, leading to a parallel arrangement of unconnected channels (mean radius $R \sim 150$ Å) aligned in a layer of thickness $\sim 30$ $\mu$m. Fully hydrogenated 8CB (Sigma-Aldrich) was confined into PSi by capillary wetting from the liquid phase under vapor pressure in a vacuum chamber and at a temperature of 60°C, well above the nematic-isotropic transition temperature [26]. Recently we have shown that the highly corrugated inner surface of the silicon pores acts as a random field, coupling to the order parameters of the confined phase [21]. This quenched disorder inhibits the nematic-to-smectic phase transition. It favors the low-temperature metastability of the smectic phase (supercooling) and prevents its correlation length from diverging: $\xi_{SmA}$ grows progressively on decreasing temperature up to a few nanometers [21,23–25] only. This structural behavior recalls the scenario describing molecular dynamics of many glass-forming liquids.

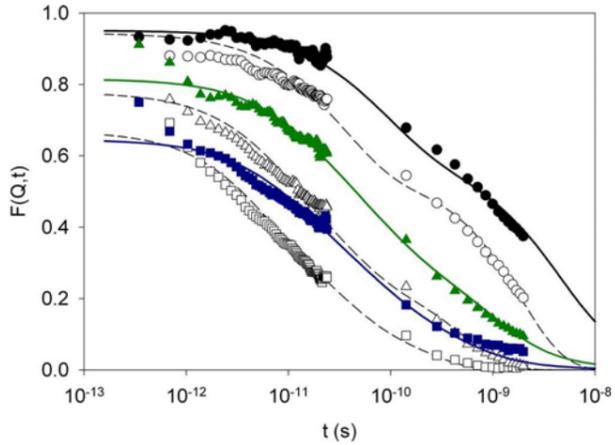

FIG. 1. (Color online) Incoherent intermediate scattering functions of bulk 8CB (open symbols) and 8CB confined in porous silicon (solid symbols) at 296 K and measured at Q=0.50 ± 0.05 (circles), 1.0±0.1 (triangles), and 1.5±0.2 Å$^{-1}$ (squares). The solid and dashed lines are best fits (see text).

Incoherent QENS experiments were carried out on the backscattering (BS) spectrometer IN16 at the Institut Laue-Langevin (ILL, Grenoble) and on the time-of-flight (TOF) spectrometer MIBEMOL at the Laboratoire Léon Brillouin (LLB, CNRS/CEA, Saclay). For IN16 the standard Si(111) setup with λ=6.271 Å results in an energy resolution of —0.9 μeV and a dynamic range of 15 μeV. The complementary resolution at MIBEMOL for λ = 6 Å was 107 μeV. Both spectrometers covered a momentum transfer (Q) range of about 0.4–1.9 Å$^{-1}$. Temperature was controlled to better than 0.1 K over a range from 10 K to 310 K (100 K to 340 K) using a cryofurnace (a cryoloop) on IN16 (on MIBEMOL). Eight wafers were stacked parallel to each other in a cylindrical aluminum cell, representing a total amount of confined LC of about 20 mg. Complete details of the experimental setup, data processing, intermediate scattering functions calculations, and model fitting are given elsewhere [11]. Figure 1 displays the incoherent intermediate scattering functions $F(Q,t)$ of nanoconfined 8CB compared to those of the bulk at the same temperature (296 K) for different momentum transfer vectors $Q$. The dynamic response of nanoconfined 8CB is globally slowed down by more than a factor 3, in agreement with previous elastic neutron-scattering measurements [26]. This slowing down is observable both in the picosecond and nanosecond regions, which are fully characterized, respectively, by uniaxial molecular rotations and translational self-diffusion relaxation mechanisms [11]. At long times, the relaxation function does not systematically decay to zero anymore on the time scale of the experiment. In that time window, long-range molecular transport dominates molecular relaxation and appears strongly affected by nanoconfinement. But also the effective correlation time $\tau_c$ associated with fast rotations increases under confinement, which is more unexpected owing to the localized character of this mode.

Despite this global slowing down, no hypothetic change in the geometry of motion, which would affect the $Q$ dependence of the elastic incoherent structure factor, can be revealed from the long time region ($t >$ 10 ps). At that point, it is also important to note that within our experiment, the condition $QR > 1$ is always fulfilled. This means that the observed slowing down cannot be due to a trivial finite-size effect, which can appear when the size R of the pore begins to compete with the typical distance probed by the neutrons [13]. On the other hand, the proximity of the solid interface may strongly affect the local dynamics of the confined molecular fluid [18–20]. Far from the pore wall, it seems reasonable to assume that the molecular dynamics results in the same incoherent scattering function $F_{bulk}(Q,t)$ as in the bulk liquid crystal. We developed a model of this function for bulk 8CB [11] including the afore-mentioned rotational and self-diffusional elementary relaxation processes. On the other hand, 8CB molecules close to the silicon wall (at a distance |r| from the pore center with R − |r| < $\xi_{wall}$) might experience a significant mobility reduction, resulting in a different local scattering function $f(r,Q,t)$. If the relaxation mechanisms remain the same as in bulk for all molecules in the pore, then they can be described by only two space-dependent parameters: the uniaxial rotation correlation time $\tau_c(r)$ and the self-diffusion coefficient $D(r)$. If their radial variation remains slow on the space scale of the experiment ($Q\xi_{wall} > 1$), the local scattering function can be expressed with the same analytical form as in the bulk:

$$f(r,Q,t) = e^{-Q^2 \langle u_{conf}^2 \rangle} F^{rot}(\tau_c(r), Q, t) e^{-D(r)Q^2 t}. \qquad (1)$$

An exponential decay $\propto \exp(-\frac{R-|r|}{\xi_{wall}})$ with distance from the pore wall is one of the simplest functional forms able to describe the inhomogeneity of $\tau_c(r)$ and $D(r)$, although alternative expressions have been proposed [15,16]. In that case and taking into account an ideal cylindrical porous geometry, the local self-diffusion coefficient of a 8CB molecule located at distance r from the pore center is written as:

$$D(r, \xi_{wall}) = (D_0 - D_\infty) \frac{\cosh\left(\frac{r}{\xi_{wall}}\right)}{\cosh\left(\frac{R}{\xi_{wall}}\right)} + D_\infty. \qquad (2)$$

Assuming the same characteristic length $\xi_{wall}$ for the uniaxial rotational mode, then $\frac{1}{\tau_c}(r)$ follows a similar law as Eq. (2). According to previous simple considerations, one has $D_\infty = D_{bulk}$ and $\frac{1}{\tau_\infty} = \frac{1}{\tau_{bulk}}$. These values are known in the studied temperature range from previous experiments on bulk 8CB [11,27]. The simplest boundary conditions are $D_0 = D(R) \to 0$ and $\frac{1}{\tau_c(R)} \to 0$ which simply means that all molecular motions of the first molecular layer anchored to the wall (except local vibrational modes) are slower than the resolution of the neutron scattering experiment. This choice is rather arbitrary and might have some limited quantitative influence on the value of $\xi_{wall}$ later derived from our work. It is, however, strongly supported by molecular dynamics simulations [16].

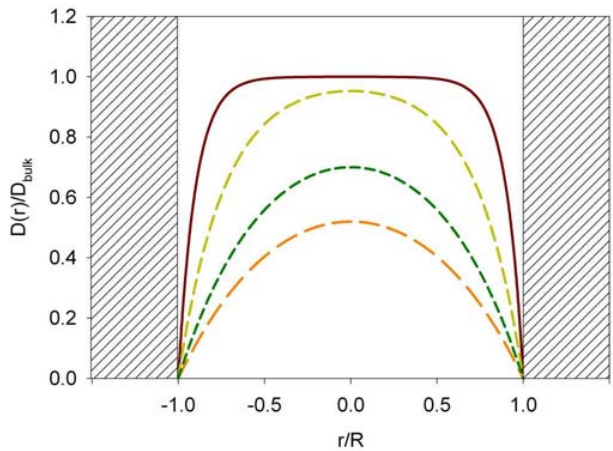

FIG. 2. (Color online) Radial dependence of the diffusion coefficient (or motion frequency) in a cylindrical pore of radius R according to Eq. (2) assuming the self-diffusion process is bulk like far from the wall. Curves from top to bottom: $\xi_{wall}/R=10$ (solid line) and $\xi_{wall}/R=0.27$, $\xi_{wall}/R=0.5$, and $\xi_{wall}/R=0.73$ (dashed lines). The dashed lines correspond to experimental values.

According to this model, a schematic view of the radial dependence of the parameters governing the molecular dynamics in the silicon pore of radius $R$ is sketched in Fig. 2. For $\xi_{wall} < R$, the distribution of these dynamic parameters is bimodal, consisting of an interfacial and a bulklike population. Conversely, for $\xi_{wall} \sim R$ the distribution is much more continuous and results in an overall slowed-down dynamics. The incoherent neutron-scattering experiment averages over the whole volume of the sample, and thus the local relaxation functions $f(r,Q,t)$ have to be averaged as well over all positions $r$ in the cylindrical pore (assuming invariance along the pore axis). The resulting spatially heterogeneous model for $F_{het}(Q,t)$ then is written as

$$F_{het}(Q,t,\xi_{wall}) = \frac{2}{R^2}\int_0^R f(r,Q,t)\,r\,dr. \quad (3)$$

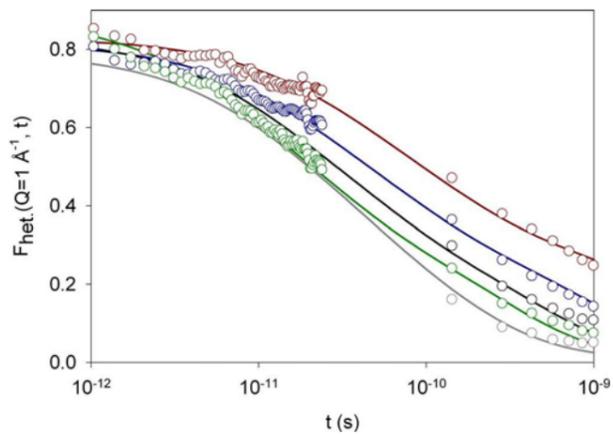

FIG. 3. (Color online) Temperature dependence of the incoherent intermediate function of 8CB confined in porous silicon measured at $Q=1$ Å$^{-1}$ and $T=280, 296, 308, 320,$ and $340$ K from top to bottom, respectively.

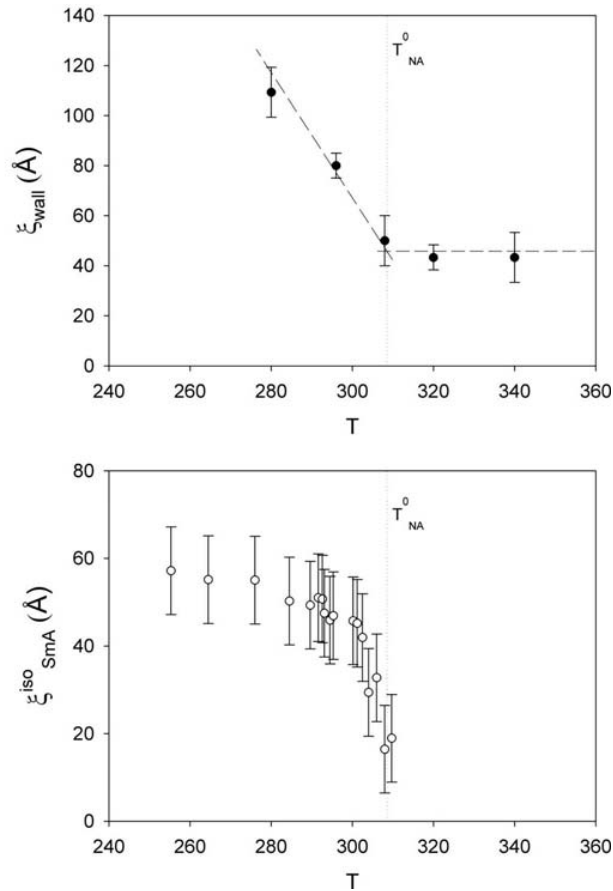

FIG. 4. Dynamic correlation length $\xi_{wall}$ (solid circles) of 8CB confined in porous compared to the average static smectic correlation length $\xi^{iso}_{SmA}$ (open circles from [21]).

We fix $R=150$ Å according to the experimental mean pore radius and assume the aforementioned blocking boundary conditions for the molecular dynamics. Within this hypothesis, there are only two free parameters left: the dynamic correlation length $\xi_{wall}$ and the effective vibrational mean square displacement $\langle u^2_{conf}\rangle$. The solid lines in Figs. 1 and 3 display best fits using such a model described by Eqs. (1)–(3). The best-fit values obtained for $\langle u^2_{conf}\rangle$ were 0.52, 0.62, 0.51, 0.47, and 0.48 Å² for temperatures $T=280, 296, 308, 320,$ and $340$ K (see Fig. 3), within a confidence interval of about 0.05 Å². This $\langle u^2_{conf}\rangle$ is found essentially constant in the studied range and hence uncorrelated with the strong temperature dependence of $\xi_{wall}$. The corresponding best fit values for $\xi_{wall}$ are displayed in Fig. 4. Both values of $\langle u^2_{conf}\rangle$ and $\xi_{wall}$ reported here are averaged over the three $Q$ moduli considered in Fig. 1. In Fig. 4, $\xi_{wall}$ is displayed in comparison with the average (isotropic) static correlation length $\xi^{iso}_{SmA} = \frac{2\xi_\perp + \xi_\parallel}{3}$ of the short-range smectic order of confined 8CB [21].

At high temperatures, where the equilibrium phases of nanoconfined 8CB are either isotropic or nematic, ~wall is found constant and its value of about 40 Å represents two to five molecular layers. The molecular dynamics seems therefore to be composed of a very slow interfacial population coexisting with a "bulklike" one in the core of the pore. On decreasing temperature, $\xi_{wall}$ increases progressively and becomes eventually comparable with the pore radius $R$ at the lowest temperature studied. The associated molecular dynamics is then rather described by a continuous distribution of correlation times and diffusion coefficients, as illustrated in Fig. 2. As this distribution is directly related to the spatial variation of $\tau_c(r)$ and $D(r)$, the correlation length $\xi_{wall}$ provides a measure of the heterogeneity of the local molecular dynamics. It is consistent with observations reported on longer time scales associated with the molecular rotation around the short axis [18–20] and to collective fluctuations of the order parameters [22,23]. It is also strongly reminiscent of the glassy character of more simple molecular supercooled liquids.

If the molecular dynamics were simply driven by the local structure, dynamic heterogeneity could be the direct consequence of a different order parameter near the wall. Indeed, at high temperatures where the LC phase is undoubtedly iso-tropic ($T>320$ K), paranematic domains can be induced at the solid-LC interface [28–30], which could result in the observed bimodal distribution of relaxation times. But this oversimplified picture cannot hold at low temperatures. It is obvious from Fig. 4 that the dynamic heterogeneity measured by $\xi_{wall}$ strongly increases as soon as short-range smectic correlations of size $\xi_{SmA}^{iso}$ appear, below the temperature where quenched disorder effects become prevailing. $\xi_{wall}$ is there much larger than the interfacial region. Most of all, in this frustrated smectic phase, the order parameters are uniform in the pore volume [21,31]. This means that the heterogeneous propagation of the interfacial slowing down is driven by more subtle cooperative effects than the homogeneous value of the order parameters only. Finally, we find that the spatial extent of this dynamic inhomogeneity follows a similar temperature dependence as that of the smectic correlations. We report in this Rapid Communication on the coupling between two correlation lengths measured independently in a confined liquid crystal. The first one ($\xi_{SmA}^{iso}$) is related to the size of the local static order, while the second ($\xi_{wall}$) is related to the memory of the interfacial interaction on the molecular dynamics. We propose that $\xi_{wall}$ should be identified as the dynamic length of cooperativity that measures the intrinsic heterogeneity of the fluid's molecular dynamics, as most usually defined for glassforming liquids. At present, no experiment has allowed a relating of the observed spatial inhomogeneous character of the molecular dynamics of nanoconfined liquids to any hypothetical supramolecular structural property. The present achievement relies on the choice of a liquid crystal, which has allowed measuring a frustrated static correlation length that escapes to the experiments on globular liquids. In the same time, all general features of the short-time dynamic correlations and intermediate scattering functions of molecular liquids are preserved. This result provides therefore a unique demonstration of how a frustrated static correlation length may couple to dynamical heterogeneities in confinement. This breakthrough is likely to contribute to the general understanding of the relation between mesoscopic order and cooperativity in many other complex systems including granular media, polymers, and glass-forming liquids.